\documentclass[final,twocolumn]{aa}
\usepackage{natbib}
\usepackage{lscape,graphicx} 
\usepackage{epsf}
\usepackage{subfig}
\usepackage{txfonts}
\normalfont

\def\rxj {1RXS J170849$-$400910}
\def\src {RXSJ1708}

\def\CXO{{\em Chandra}}
\def\RXTE{{\em Rossi-XTE}}
\def\ergscm2{\rm erg\,cm^{-2}\,s^{-1}}

\newcommand {\rc}{\rm }
\begin{document}

\title{Linking the X--ray timing and spectral properties of the glitching AXP
\rxj\ }

\author{G.L. Israel\inst{1}, D. G\"otz\inst{2}, S. Zane\inst{3}, S.
Dall'Osso\inst{1}, N.
Rea\inst{4} and L. Stella\inst{1}}

\institute{INAF - Osservatorio Astronomico di Roma, Via Frascati 33, 00040 Monte
Porzio Catone, Italy \and CEA Saclay, DSM/DAPNIA/Service d'Astrophysique,
  F-91191, Gif sur Yvette, France  \and Mullard Space Science Laboratory
University College of London
Holmbury St Mary, Dorking, Surrey, RH5 6NT,  UK \and SRON - Netherlands
Institute for Space
Research, Sorbonnelaan 
2, 3584 CA, Utrecht, the Netherlands}

\offprints{gianluca@mporzio.astro.it}

\date{Received / Accepted}

\abstract
{}{Previous studies of the X-ray flux and spectral properties of \rxj\ showed
a possible correlation with the spin glitches that occurred in 1999 and
2001. However, due to the sparseness of spectral measurements and the paucity of
detected glitches, no firm conclusion could be drawn.}
{We retrieved and analysed archival \RXTE\
pointings of \rxj\ covering the time interval between January 2003 and June 2006
and carried out a detailed timing analysis with phase fitting
techniques.}{We detected two large glitches ($\frac{\Delta
\nu}{\nu}$ of 1.2 and 2.1$ \times 10^{-6}$) that occurred in January and June
2005.
The occurrence times of these glitches are in agreement with the
predictions made in our previous studies.
% and based on the flux and spectral parameters
%evolution. 
This finding {{\rc strongly suggests}} a connection between the
flux, spectral and timing properties of \rxj.}{}


\authorrunning{Israel et al.}
\titlerunning{\rxj\ spectral/timing correlations}

\maketitle

\section{Introduction}
\rxj\ (hereafter \src) is one of the anomalous X-ray pulsars (AXPs), a small
group of 
peculiar neutron stars (NSs) that 
are currently believed to have super-strong magnetic fields, $B\sim 
10^{14}-10^{15}$~G, hence dubbed ``magnetars'' \cite[though other 
possible models
are not completely ruled out by observations; for a recent review see][]
{wt06}. \src\ was first discovered by ROSAT
\citep{vog96}, while $\sim$\,11\,s coherent pulsations were  
detected with ASCA \citep{sug97}. Early measurements suggested
that it was a fairly stable rotator, with a spin-down rate
of $\sim2\times10^{-11}$s\,s$^{-1}$, and a soft spectrum \citep{gia99, gia01}. 
Events of sudden spin--up (glitches) with very different post-glitch recovery
were detected in  \src\ by \RXTE\ in 1999 and 2001 \citep{ka00,simone03,kg03}.
The rather short interglitch time makes this AXP a frequent glitcher among
neutron stars. 

Recently, \cite{na05} noticed that
the long term (over 5 years) variations in the source X-ray flux and spectral
hardness are correlated, with both quantities reaching a maximum  close to the
epochs of the two glitches in 1999 and 2001. The addition of new data obtained
from  {\em Chandra} (2004) and {\em Swift} observations confirmed the 
flux-hardening correlation \citep{ca07}, extending its validity
to hard X-rays \citep{diego07}.  
Following these studies, \cite{silvia07} and \cite{na07} suggested that the
long-term variations may have a cyclic behavior with a recurrence time of
$\approx 5$-10yrs, possibly due to a periodic twisting/untwisting of the star
magnetosphere \citep{bt07}. Correspondingly, the source was expected to re-enter
into a glitching active phase during 2005-2006, close to the latest maximum in
the source flux.

Thus, we analyzed the \RXTE\ archival data spanning the latest 3.5
years and performed a phase coherent analysis of the pulse arrival times in
a search for new glitches. We detected two large glitches occurring on 
January 2005 and May 2005, in agreement with the prediction of 
\cite{silvia07} and \cite{na07}  and correlating with the flux
and spectral evolution.

\section{\RXTE\ observations and timing analysis}

We analyzed 204 \RXTE\ archival observations of
\src\footnote{ftp://legacy.gsfc.nasa.gov/xte/data/archive}.  These span
from 2003 January 5th to 2006 June 3rd.  
We restricted our analysis to the PCA instrument \citep{pca96} which was
operated in good Xenon data mode with a time
resolution of 1 $\mu$s and 256 energy bins between 2 and 120 keV. Raw data
were reduced using the {\tt ftools} v6.2, provided
by the High Energy Astrophysics Science Archive Research
Center\footnote{
http://heasarc.gsfc.nasa.gov/docs/software/ftools/ftools\_menu.html}.
The events were extracted in the 2.5--16 keV energy range, and binned into light
curves of 0.125 s resolution. Photon arrival time correction to the barycentric
dynamical time (TDB) was applied by using {\tt fxbary} and the
($\sim$1$''$ accurate) source position provided by \cite{gia03}.

\begin{figure}[tb]
 \centering
 \hspace{-0.5cm}
  \includegraphics[angle=-90,width=0.5\textwidth]{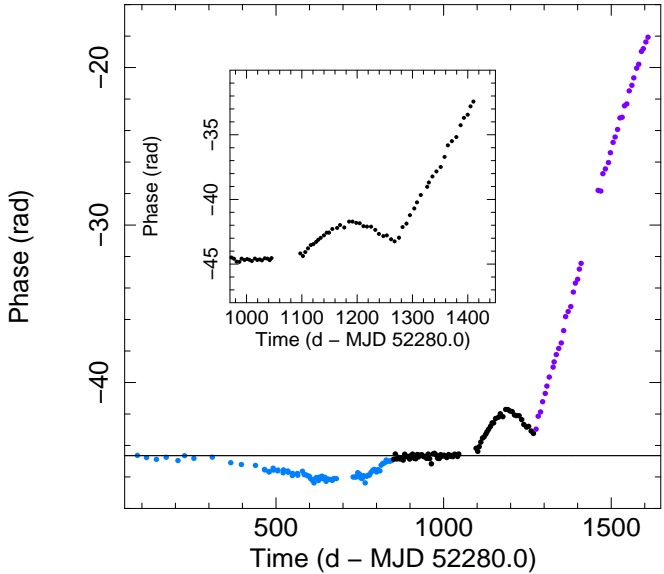}
 \caption{\RXTE\ time residuals for the time interval January 2003 -- June 2006
after subtraction of the phase-coherent P-\.P solution {\rc inferred by
\cite{simone03}}. The inset shows the time interval over which we
detected the two glitches.}
 \label{fig:his_fin}
\end{figure}

A phase--coherent timing solution was first obtained using a
relatively long (29ks) archival \CXO\ observations carried out on 2004 July 3rd 
\citep[starting time at MJD 53189.151489 TDB; for details on the data reduction
of this dataset see][]{no05sgr}. This provided a
sufficiently accurate period determination, P$=11.00231(3)$s, such that no pulse
cycle was missed when extrapolating this value to the epoch of the closest
\RXTE\ pointing (2004 July 1st). A phase-coherent timing
solution was inferred in the time interval between 2004
May 1st and November 16th corresponding to  
$\nu$=\,0.090890035(1) Hz and $\dot{\nu}$=\,-1.5884(14) (epoch 53819.0 MJD;
1$\sigma$ c.l. are reported). 

\begin{figure}[tb]
 \centering
 \hspace{0cm}
  \includegraphics[angle=-90,width=0.44\textwidth]{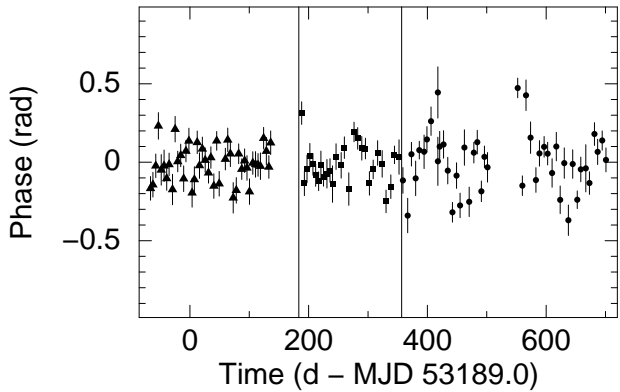}
 \caption{Time residuals of \RXTE\ \src\ observations from June 2004 to June
2006 after subtraction of pre-- and post--glitch model N.1 and the polynomial
post--glitch model N.2. Vertical lines indicate the assumed/inferred epochs of
the glitches.}
 \label{fig:residuals}
\end{figure}

The inclusion of the 2005 and 2006 datasets showed large disagreement with the
above inferred timing solution, with two evident "jumps" in phase both  marking 
a decrease in the period value (see blue and dark
violet filled circles in Figure \ref{fig:his_fin}), strongly
suggesting that two further glitches occurred at the end of 2004 / begin 2005
(MJD 53370) and in May 2005 (MJD 53550). No signature for similarly large
glitches was instead found in the datasets spanning between January 2003 and
June 2004. {\rm The phases reported in Figure
\ref{fig:his_fin} are obtained using the \cite{simone03} 2001 post-glitch
solution (horizontal solid line) which is nearly coincident to the phase
coherent solution inferred above.}
Thus, we  applied a detailed timing analysis to phase
residuals were the two large glitches occurred. A more exhaustive study
and modelling of the whole phase and glitch history of \src\ will
be presented elsewhere \citep{simo07}. Following the analysis
scheme outlined in \cite{simone03} we inferred
the main parameters of the two detected large glitches (see Table
\ref{glitches}). 
\begin{table}
\begin{center}
\caption{Measured parameters for the two glitches detected in the \RXTE\
 data of \src. 1$\sigma$ errors in the last 
digit are quoted in parenthesis.}
\label{glitches}
\begin{tabular}{lcc}
\hline
\hline
Spin Parameter & post-glitch N.1 &  post-glitch N.2\\
\hline
Epoch (MJD) & 53372(2) &  53546.0(8) \\
$\nu$ (Hz) & 0.090887638(16)  &  0.09088524(20) \\
$\dot{\nu}$ ($\times 10^{-13} s^{-1}$) & -1.700(4)  & 
-1.536(7)\\
$\ddot{\nu}\times 10^{-22} s^{-3}$ & $<$ 4.7  &  -3.78(34) \\
MJD range & 53372-53545 & 53546-53889  \\
N. datapoints & 29 & 45 \\
r.m.s. (s) & 0.26 & 0.39   \\
$\Delta \nu / \nu (\times 10^{-6})$ & 1.18(3) & 2.08(5) \\
$\Delta \dot{\nu} / \dot{\nu} \times 10^{-2}$ & 7.0(3) & -10.35(34)\\
\hline
\end{tabular}
\end{center}
\end{table}
Figure \ref{fig:residuals} shows the phase residuals after subtraction of
the timing solutions pre-- and post--glitch N.1, and post--glitch N.2.
Both the newly detected glitches reveal large jumps in the spindown rate, 
$\Delta \dot \nu/ \dot{\nu} \sim 7 \times 10^{-2}$ and $\Delta \dot \nu / 
\dot{\nu} \sim - 0.1$, among the largest ever observed in 
glitches that lack a significant short-term exponential recovery. Remarkably,
they have opposite signs: the second glitch has cancelled the effect of the
previous increase in $\dot{\nu}$ and, actually, somewhat overshot it. 

The jump in spin frequency after the first glitch appears have been
recovered  in $\sim$ 120~d.
The upper limit on $\ddot{\nu}$ after the first glitch implies that the jump in
$\dot{\nu}$ could have been recovered, if at all, only on a much longer
timescale and 
not until the second glitch occurred, $\sim$ 175~d after the first one.
At the second glitch, an even larger spin-up occurred, accompanied by a 
significant flattening of the spindown trend. Thus, the spin up started 
with a sudden increase and then it slowly continued. A 
large and negative second derivative is found to be highly significant in the 
fits, which corresponds to a long-term recovery of $\Delta \dot{\nu}$ on 
a timescale $\simeq (503 \pm 48)$~d. We note that the
$\ddot{\nu}$ term detected after the second glitch brings $\dot{\nu}$
back to the value it had before the first glitch in $\sim (159 \pm 26)$~d, and
to steeper values afterwards. The extrapolated spin frequency at 500~d from 
the second glitch gives an additional spin-up $\sim (5.7 \pm 0.5) \times 
10^{-7}$ s$^{-1}$, while at the 
same epoch $\dot \nu$ returned to the value $\dot{\nu} (\Delta t =500 
{\rm d}) \simeq - 
(1.70 \pm 0.02) \times 10^{-13}$ s$^{-2}$ that it had after the first 
glitch.

\section{Discussion and conclusions}

\src\ experienced two new glitches, both with a large
fractional amplitude, $\Delta \nu/\nu \sim 1.2\times 10^{-6}$ and $ 2.1 \times
10^{-6}$ respectively, comparable to the so-called giant glitches observed from
Vela and to the glitch previously detected from this source in May 2001
\citep{simone03, kg03}. This result strongly suggests that giant glitches
seem to be the rule for this source.

It has been noticed \citep[see][\, for details]{simone03,wang00} that 
most glitch models are difficult to reconcile with observations of a growing
number of glitches. In particular, all models requiring a catastrophic (\textit
{i.e.} widespread) unpinning of crustal superfluid \citep[such as the vortex
creep
models, see][]{alp84, alp93, alp89} seem less promising, since they
require simple correlations between the amplitudes and recovery timescales of
glitches in a single source as well as an approximately constant Q-value (where
Q is the recovery fraction, see \cite{simone03}), which
is found not to be the case, in general, for radiopulsars. 
Although these problems can be partially solved by invoking local unpinning, 
this appears to be a rather ad hoc assumption \citep{jones02}. Furthermore, it
has been recently pointed out that the observation of long-period precession in
a few pulsars is incompatible with pinning of crustal vortices \citep{link06}.

Models where the internal angular momentum reservoir is in the NS core have 
some advantages with respect to the above points and can in principle 
explain a wider range of glitch properties \cite[e.g.see][\, for a more
complete discussion]{simone03}. In this respect, it is worth noticing the
unusually large jumps in the spindown rate of the two new events.  
A $\sim$ 7\% increase found after the first glitch would rule out the crustal
superfluid as a momentum reservoir: the \textit{core} superfluid 
- and a remarkable fraction of it - would necessarily be required to have been 
involved in the glitch. On the other hand, the large flattening that follows the
second glitch is also very peculiar \cite[although a residual steepening
of the spindown trend, not recovered promptly, was also found in the May 
2001 glitch from this same source,][]{simone03}. 
In general, it is not easy to explain long-term offsets from the secular
spindown trend in glitch models based solely on crustal superfluid instabilities
\citep{Link92}, since these require permanent decoupling of some fraction of
the internal superfluid from the bulk of the NS mass. More extreme difficulties
are posed by the slowly recovered decrease in the spindown rate found in 
the second glitch: this apparently implies either a re-coupling of an internal 
component that was previously decoupled from the crust, or a corresponding 
decrease in the spindown torque.

A potential argument against core-based models arises when using
$\dot{\nu}_{gl}$ (the average spin--up rate caused by glitches) to estimate the
fractional moment of inertia of the internal reservoir of angular momentum
\citep{lyne00,simone03}. The parameters of the three large
glitches detected so far from \src\ yield  $\dot{\nu}_{gl} \simeq 2.65 \times
10^{-15}$ s$^{-2}$. Since the ratio  $\dot{\nu}_{gl}/\dot{\nu} \leq 
I_{res}/I_c$, one derives $I_{res}/I_c \geq 0.017$ where $I_{res}$
and $I_c$ are the reservoir and stellar moments of inertia. This is
comparable with the value inferred typically in glitching radio pulsars, and
consistent with the crustal superfluid being the momentum 
reservoir \citep{lyne00}. However, we notice 
that the equality in the above formula obtains in the ideal case where the
internal reservoir is completely decoupled from the crust. 
In the presence of coupling this component would also spin--down  in
between
glitches at a rate $\dot{\nu}_{res}<\dot{\nu}$. Therefore, the following
equality,
$\dot{\nu}_{gl}/(\dot{\nu} - \dot{\nu}_{res}) = I_{res}/I_c$ should be
considered instead. If glitches are produced
once a critical rotational lag $\nu_{cr}$ is reached, then 
$\dot{\nu} - \dot{\nu}_{res} \sim \nu_ {cr}/ t_g$, where $t_g$ is the
typical interglitch time. If we assume $I_{res} \simeq 0.1 I_c$, as 
required to explain the changes in $\dot{\nu}$ found for the two new 
glitches,  the measured $\dot{\nu}_{gl}$ and a typical 
interglitch time $\leq$2ys, we find $\nu_{cr} \leq 1.6 \times 10^{-6}$ 
s$^{-1}$. Whether such a low critical lag is meaningful on physical grounds
is matter for future investigation \cite[see][for similar 
conclusions]{simone03}.
\begin{figure}[tb]
 \centering
  \includegraphics[angle=-90,width=0.33\textwidth]{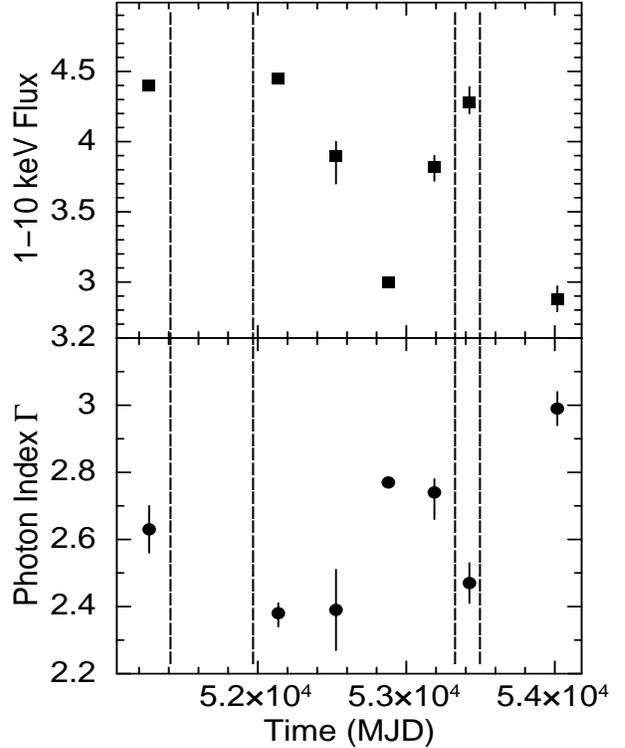}
\caption{Long-term spectral history of \src\ showing the correlated 
hardness--intensity variation. All reported fluxes are absorbed (with 
$N_{H} = 1.36\times10^{22}$\,cm$^{-2}$ for all the spectra) and in the 
1-10\,keV energy range (see \cite{diego07} for further details). Vertical
dashed lines mark the glitches detected in 1999, 2001, January and May 2005,
from left to right respectively. {\rc Flux cross calibrations among instruments
are less than 15\% (note that the latest two measurements are both inferred from
Swift XRT and implies a flux variability of more than 25\%.}
 }
 \label{fig:spec_time}
\end{figure}
An alternative explanation, in terms of starquakes and subsequent 
movements of cracked platelets, has been proposed for at least the large
glitches of radio pulsars \citep{ruderman91,ruderman98}. 
Quakes may be caused by the stress acting on the crust following the 
interaction in the core between superfluid neutrons and magnetic flux tubes
threading the crust.
Alternatively, the crustal strain that is accumulated during the growing of a
twist in a magnetar's magnetosphere may represent a trigger in AXPs. Only
the moving sector and its surroundings are affected, leaving the rest of the
star mainly unperturbed \citep{jones02}. The ``local nature'' of starquake
models make them promising for reproducing the large variety of
properties found in AXP glitches \citep{simone03}. 

In  \cite{na05}, we proposed that the observed correlation between the X-ray 
flux and the spectral hardness may be explained if the evolution is regulated 
by the change of a ``twist'' in the magnetosphere \cite[see also][]{tlk02}. 
The evolving magnetic field is expected to fracture the crust at intervals,
eventually causing an increased activity and large amplitude glitches. We found
that observations collected until 2003 were consistent with a scenario in which
the twist angle was steadily increasing before the glitch epochs, culminating
with glitches and a period of increased timing noise, and then decreasing,
leading to a smaller flux and a softer spectrum {\rc (see Figure
\ref{fig:spec_time})}. 
Interestingly, the same model provides a natural explanation for the new period
of glitching activity, that was foreseen in our previous papers. 
Nonetheless, we emphasise that while we do expect glitching activity 
corresponding of an increasing stress of the crust caused by a growing twist,
glitches might also occur  outside these epochs, in particular if,
as stressed before, glitches with different properties (such as amplitude and
recovery) may reflect a difference in triggering mechanism. 
{\rc We tried to estimate a reliable false alarm probability for 
the observed glitches to occur by chance close to the flux maxima displayed by
observations, by assuming a uniform probability of flux values at glitch
epochs. The false alarm probability depends on the maximum assumed flux range
(2.5$\times$10$^{-11}\ergscm2$, 60\% more than that observed; see Figure
\ref{fig:spec_time}) and the given flux threshold value
(4.2$\times$10$^{-11}\ergscm2$) above which glitches occur\footnote{\rc We
implicitly assumed a uniform distribution for glitch epochs (which is likely
not true).}. 
The above, somewhat conservative assumptions give a false
alarm probability of 3\%.
Based on the above considerations we
believe that the proposed link between timing and spectral properties of \src\
is rather good. However, given the poor knowledge of the true flux history and
the not statistically independent nature of glitches, we believe that only 
future observations will be able to unambiguously confirm the link suggested by
observations. Nonetheless, when the 1999 glitch
is also added to the above considerations, the false alarm probability decreases
to 1\%.} Thus, if such correlated long-term variation
will be further confirmed and discovered in other AXPs/SGRs, its X-ray
monitoring might become an excellent tool to gather a deeper understanding of
magnetars and, more generally, neutron star glitches.

{\rc Finally, we note that \cite{rim07} submitted a paper on an independent
discovery of glitches from \src\ in the same dataset (after our original
report at the Seattle Workshop on 27th
June 2007\footnote{http://www.astroscu.unam.mx/neutrones/INT/Workshop.html}) :
they considered our glitch N.\,1 as a candidate rather than a true glitch. A
detailed comparison between their results and ours is indeed premature and
beyond the aims of the present work \citep{simo07}. Nonetheless, we note that
a source of discrepancy is due to the use by \cite{rim07} of high-order
frequency derivatives (used to identify glitches) in the presence of gaps in
the phase-series. It is indeed not surprising that two out of the three
candidate glitches reported by \cite{rim07} are found when data gaps are also
present.}

\begin{acknowledgements}
This work is partially supported at OAR through Agenzia Spaziale Italiana (ASI),
Ministero  dell'Istruzione, Universit\`a e Ricerca Scientifica e Tecnologica
(MIUR -- COFIN), and Istituto Nazionale di Astrofisica (INAF) grants. We
acknowledge financial contribution from contract ASI-INAF I/023/05/0. 
SZ acknowledges support from a STFC (ex-PPARC) AF. DG acknowledges financial
support from the
French Space Agency (CNES).
\end{acknowledgements}

\end{document}